\PassOptionsToPackage{table}{xcolor}
\documentclass[manuscript,screen, acmsmall]{acmart}
\usepackage{subcaption}
\usepackage{multirow}
\usepackage{tcolorbox}
\usepackage{threeparttable}
\usepackage{tikz}
\usepackage{pgfplots}
\usepackage{wrapfig}
\definecolor{customorange}{HTML}{ffa602}
\definecolor{customblue}{HTML}{6394ee}
\definecolor{lightorange}{HTML}{ffd580} % softer orange
\definecolor{lightblue}{HTML}{a7c7f2}   % softer blue
\definecolor{darkred}{RGB}{139,0,0}
\definecolor{darkblue}{RGB}{0,0,139}
\AtBeginDocument{%
  }

\acmISBN{978-1-4503-XXXX-X/2018/06}

\definecolor{VibrantBlue}{rgb}{0.0, 0.2, 1.0}

\definecolor{ForestGreen}{rgb}{0.13, 0.55, 0.13}

\begin{document}

% \title{On Agreement Dynamics: How Model Scale and Task Design Shape Persona-Based Content Moderation Systems}
\title{LLM-Generated Ads: \\From Personalization Parity to Persuasion Superiority}

\author{Elyas Meguellati}
\orcid{0009-0003-4982-9565}
\affiliation{%
  \institution{The University of Queensland}
  \city{Brisbane}
  \country{Australia}}
\email{m.meguellati@uq.edu.au}

\author{Stefano Civelli}
\orcid{0009-0003-4982-9565}
\affiliation{%
  \institution{The University of Queensland}
  \city{Brisbane}
  \country{Australia}}

\author{Lei Han}
\orcid{0009-0003-3657-9229} 
\affiliation{%
  \institution{The University of Queensland}
  \city{Brisbane}
  \country{Australia}
}

\author{Abraham Bernstein}
\orcid{0009-0003-4982-9565}
\affiliation{%
  \institution{The University of Zurich}
  \city{Zurich}
  \country{Switzerland}}

\author{Shazia Sadiq}
\orcid{} 
\affiliation{%
  \institution{The University of Queensland}
  \city{Brisbane}
  \country{Australia}
}

\author{Gianluca Demartini}
\orcid{0000-0002-7311-3693} 
\affiliation{%
  \institution{The University of Queensland}
  \city{Brisbane}
  \country{Australia}
}

\renewcommand{\shortauthors}{Meguellati et al.}

\begin{abstract}
As large language models (LLMs) become increasingly capable of generating persuasive content, understanding their effectiveness across different advertising strategies becomes critical. This paper presents a two-part investigation examining LLM-generated advertising through complementary lenses: (1) personality-based and (2) psychological persuasion principles.
In our first study (n=400), we tested whether LLMs could generate personalized advertisements tailored to specific personality traits (openness and neuroticism) and how their performance compared to human experts. Results showed that LLM-generated ads achieved statistical parity with human-written ads (51.1\% vs. 48.9\%, $p > 0.05$), with no significant performance differences for matched personalities.
Building on these insights, our second study (n=800) shifted focus from individual personalization to universal persuasion, testing LLM performance across four foundational psychological principles: authority, consensus, cognition, and scarcity. AI-generated ads significantly outperformed human-created content, achieving a 59.1\% preference rate (vs. 40.9\%, p<0.001), with the strongest performance in authority (63.0\%) and consensus (62.5\%) appeals. Qualitative analysis revealed AI's advantage stems from crafting more sophisticated, aspirational messages and achieving superior visual-narrative coherence. Critically, this quality advantage proved robust: even after applying a 21.2 percentage point detection penalty when participants correctly identified AI-origin, AI ads still outperformed human ads, and 29.4\% of participants chose AI content despite knowing its origin. These findings demonstrate LLMs' evolution from parity in personalization to superiority in persuasive storytelling, with significant implications for advertising practice given LLMs' near-zero marginal cost and time requirements compared to human experts.
\end{abstract}

\maketitle

\section{INTRODUCTION}

The rapid emergence of Generative AI (GenAI) is reshaping how marketing content is produced and deployed, with early evidence that such systems can materially affect creative tasks and workflows in practice \cite{Grewal2025GenAI}. In advertising, this shift intersects with a longstanding, resource-intensive creative process that relies on expert iteration and collaboration within agencies \cite{Turnbull2017CreativeProcess}. A central question follows: \textit{Can large language models (LLMs) achieve parity with or surpass human experts in the generation of persuasive communications that effectively engage consumers?}

Understanding LLM effectiveness in advertising requires engaging with two pillars of persuasive communication. First is \emph{personalization}: tailoring messages to individual differences, for which the Big Five framework provides a widely used lens \cite{Goldberg1993BigFive}. Personality-targeted advertising has shown that aligning message framing with traits can increase effectiveness at scale \cite{Matz2017PsychTargeting,Winter2021TraitAds}. Second is \emph{persuasion}: leveraging broadly applicable influence mechanisms that shape judgments independent of narrow targeting. Decades of work document the roles of source credibility (authority) \cite{Pornpitakpan2004Credibility}, social norms and consensus \cite{Cialdini1990Norms}, processing fluency and cognition \cite{Alter2009Fluency}, and scarcity signals \cite{Lynn1991Scarcity} in driving compliance and choice.

A further consideration is whether the \emph{origin} of content (human vs.\ AI) changes consumer response. Behavioral research shows both “algorithm aversion,” where people discount algorithmic outputs after observing errors \cite{Dietvorst2015AlgAversion}, and “algorithm appreciation,” where people sometimes prefer algorithmic to human advice \cite{Logg2019AlgAppreciation}. These mixed priors make advertising a particularly revealing domain for testing whether high-quality execution by LLMs can overcome origin-based biases.

This paper presents a two-part investigation that evaluates LLM-generated advertising across both the personalization and persuasion paradigms. In Study~1, we test whether LLMs can implement trait-based personalization (openness, neuroticism) with effectiveness on par with human experts when audiences are trait-matched. Study~2 shifts to universal persuasion, comparing LLM- and human-crafted ads built around authority, consensus, cognition, and scarcity principles. Across both studies, we combine quantitative preference data with qualitative analysis to examine which ads participants prefer, why they prefer them, and whether origin (human vs.\ AI) matters.

Our findings reveal a progression. In Study~1, LLM-generated ads achieve statistical parity with human experts for matched personalities. In Study~2, when applying universal persuasion principles, LLM-generated ads significantly outperform human-created content (59.1\% vs.\ 40.9\%, $p<.001$), and this superiority remains robust even when participants correctly identify content as AI-generated. Qualitative analyses suggest that LLMs deliver more aspirational messaging and stronger visual–narrative coherence, indicating a shift in which execution quality—rather than human authorship—may increasingly determine advertising effectiveness.

\subsection*{Contributions}

This paper makes two primary contributions:

\begin{enumerate}
\item \textbf{Empirical evidence of LLM advertising effectiveness across paradigms:} We provide the first systematic comparison of LLM-generated versus human-created advertising across both personalization and persuasion strategies, revealing conditions under which LLMs achieve parity versus superiority.\footnote{This article is an extended version of our short paper published in the Companion Proceedings of the ACM Web Conference 2024 \cite{meguellati2024good}. Compared to the preliminary version which focused solely on personality-based text generation (Study 1), this work introduces a completely new experiment evaluating multimodal persuasion principles (Study 2), adds a detailed qualitative analysis of user preference drivers (Appendix A), and provides new demographic insights regarding AI Preference and bias.}

\item \textbf{Evidence of quality transcending origin:} Through mixed-methods analysis, we show that high-quality execution can overcome bias against AI-identified content. Specifically, we find that even after applying a 21.2 percentage point detection penalty, AI ads still outperform human ads, and 29.4\% of participants chose AI content despite knowing its origin—challenging assumptions about the necessity of human authenticity in persuasive communication.
\end{enumerate}

% The remainder of this paper proceeds as follows. Section~2 reviews related work on personality-based advertising, persuasion principles, and LLM capabilities. Section~3 presents Study~1 (trait-based personalization). Section~4 presents Study~2 (persuasion principles). Section~5 integrates findings across studies. Section~6 discusses limitations and ethics, and Section~7 concludes.

\section{RELATED WORK}

\subsection{Personalization in Advertising}
Personalizing ads based on users' personality traits has gained significant attention in recent years due to its potential to enhance engagement and effectiveness. \citet{youyou2015computer} demonstrated the potential of computer-based models in achieving more accurate personality judgments using Facebook Likes, offering insights into the future of social-cognitive activities in various domains. \citet{kosinski2013private} highlighted that personal traits and attributes can be predicted from digital records of human behavior, such as Facebook Likes, with a high degree of accuracy, raising both opportunities and concerns about privacy.

\citet{matz2017psychological} discussed the effectiveness of psychological targeting for digital mass persuasion and found that substantial changes in behavior were observed when tailoring persuasive appeals to the person's psychological traits. Similarly, \citet{zarouali2022using} investigated the effectiveness of political microtargeting (PMT) by tailoring political ads to citizens' personality traits and found that customizing political ads to match citizens' personality traits results in greater persuasion.

\citet{chen2015making} explored the relationship between personality traits and ad engagement in social media advertising and demonstrated the effectiveness of using derived personality traits for social media ad targeting. \citet{winter2021effects} investigated the effects of microtargeting on consumer persuasion in the context of social media advertising and found that ads tailored to specific persuasive strategies and personality traits showed higher engagement intentions.

\citet{shumanov2022using} their approach involves utilizing contextual information to determine consumer personality traits, they demonstrated that aligning consumer personality with corresponding advertising messages can result in more successful persuasion. \citet{ning2019personet} proposed a Friend Recommendation System (FRS) that leverages the five traits and hybrid filtering, taking into account both traits and harmony ratings for friend suggestions, showing potential for enhancing personalized user experiences on social media platforms.

\citet{lee2018advertising} examined the impact of social media advertising content on customer engagement, specifically focusing on emotional, humorous, and informative content. They found that brand personality, particularly philanthropic gestures, increases customer engagement, while informative content does not perform well. \citet{ribeiro2019microtargeting} provided insights on the abuse and exploitation of Facebook's targeted advertising infrastructure for political gain and suggested ways to redesign the infrastructure to prevent such abuse, which can have implications for ad personalization and targeting strategies.

In addition to the previously mentioned studies, \citet{hirsh2012} investigated the effectiveness of tailoring ads based on individuals' personality traits and found that ads designed to match the recipient's psychological profile were more effective in changing consumer attitudes . Moreover, \citet{winter2021effects} demonstrated that personality-targeted ads can increase user engagement in social media advertising compared to non-targeted ads, indicating the potential for personality-based targeting to improve the overall effectiveness of advertising campaigns.

The literature on personalizing ads based on users' personality traits has demonstrated the potential of leveraging digital records of human behavior and computer-based models for more accurate personality judgments and targeting. However, it also raises concerns about privacy and the potential for abuse. These studies highlight the importance of responsible ad targeting, transparency, and control over users' information, as well as the need to carefully consider the ethical implications of ad personalization \citep{kosinski2013private,ribeiro2019microtargeting}. Furthermore, the findings suggest that trait-based personalization can enhance the persuasiveness of advertising messages by increasing the relevance and resonance of the content with the target audience \citep{chen2015making,winter2021effects,shumanov2022using}.

In addition to what has been stated, studies have been conducted to investigate the interactions between personality traits and factors such as emotional, humorous, and informative content, to optimize ad performance and engagement \citep{lee2018advertising}. These studies have provided insights into the complex interplay between personality traits and advertising content, including the impact of different types of content on customer engagement and the role of brand personality in influencing consumer behavior.

However, the ethical implications of using personal data for targeting purposes have also been a concern in the field of personalized advertising. Researchers have examined the need to protect users' privacy and prevent the abuse of targeted advertising infrastructure, particularly in sensitive domains like political advertising \citep{zarouali2022using,ribeiro2019microtargeting}. These studies have emphasized on the importance of responsible ad targeting and the challenges of balancing effective personalization with ethical considerations.

Having explored the related work on personalization in advertising, the next subsection will discuss the use of large language models (LLMs) and their capabilities, further expanding the scope of the literature review and providing a comprehensive foundation for the study of personalized ads creation using LLMs.

\subsection{The Emergence of LLMs}
Significant progress has been made in LLMs, with increased scale and data volume contributing to enhanced performance in a variety of tasks, in accordance with the established scaling law \citep{kaplan2020scaling}. Models such as GPT-3 (175B parameters) and PaLM (540B parameters) demonstrate capabilities and behaviors distinct from smaller models like BERT (330M parameters) and GPT-2 (1.5B parameters), thanks to their emergent abilities \citep{wei2022emergent}. LLMs' capacity to handle complex tasks, including few-shot learning via in-context understanding, is exemplified by GPT-3 \citep{shanahan2022talking,wei2022chain,hoffmann2022training,taylor2022galactica}.

The advent of innovative LLMs, such as ChatGPT, and the anticipation surrounding GPT-4 have led to debates within the AI community about the possibility of artificial general intelligence (AGI), with some arguing that GPT-4 could be the first iteration of an AGI system \citep{bubeck2023sparks}.

LLMs' impact extends beyond Natural Language Processing (NLP) and encompasses fields such as Information Retrieval (IR) and Computer Vision (CV). In NLP, LLMs serve as all-purpose language task solvers. In IR, AI chatbots like ChatGPT challenge conventional search engines, while platforms like New Bing employ LLMs to improve search results \citep{huang2023language,cao2023comprehensive,driess2023palm,wu2023visual}. In CV, vision-language models such as GPT-4 are being developed to enable multimodal dialogues \citep{zhao2023survey}.

However, while the progress and implications of LLMs are remarkable, there are still challenges in understanding the underlying principles that contribute to their superior abilities \citep{wei2022emergent,fu2022does}. Questions persist regarding the emergence of these capabilities and the factors that drive this phenomenon. Additionally, the practical aspects of training proficient LLMs, including computational resource requirements and ensuring alignment with human values and preferences, necessitate further research and development\footnote{https://cdn.openai.com/papers/gpt-4.pdf}.
%https://arxiv.org/abs/2303.08774
As the field continues to advance, continued exploration of LLMs and addressing these challenges will pave the way for unlocking their full potential in various domains. Our study on utilizing LLMs for generating personalized ads complements existing work by enabling automated targeting at scale, leveraging their power to revolutionize the delivery of tailored advertisements. 

\subsection{LLMs for Personalized Persuasion}

LLMs have rapidly advanced the study and practice of personalized persuasion across various application areas such as public policy, health, advertising, and dialogic settings. Empirically, several studies demonstrate that LLMs can generate messages that are at least as persuasive as typical human baselines, and sometimes more so, especially when tailored to audience characteristics. For instance, \citet{matz2024potential} show across four studies that ChatGPT-crafted messages aligned to recipients’ psychological profiles outperform non-personalized alternatives, underscoring the scalability of microtargeted generation. In the political domain, \citet{bai2025llm} report that LLM-authored messages can meaningfully shift attitudes on policy issues at magnitudes comparable to human-written texts, albeit via distinct stylistic pathways (e.g., perceived factuality and logical tone for AI vs.\ originality for humans). Moving to interactive debate, \citet{salvi2025conversational} find that GPT-4, when given access to demographic cues for personalization, ``outpersuades'' a human interlocutor in roughly two-thirds of trials, highlighting the potency of personalization in dialogue.

At the same time, evidence is mixed on the incremental value of fine-grained microtargeting relative to strong generic messaging. Using GPT-3 to instantiate multiple persuasive strategies (including microtargeting and interactive elaboration), \citet{argyle2025testing} find significant attitude change overall from AI-generated content but only modest gains from personalization beyond a well-crafted generic message. In health communication, \citet{karinshak2023working} show that GPT-3 can produce pro-vaccination messages rated more effective than official guidance, while also revealing that source labeling (AI vs.\ human) shapes reception—pointing to the importance of disclosure and perceived agency in persuasive outcomes.

Alongside efficacy, emerging work flags ethical and safety risks when LLMs are tasked with influence. \citet{liu2025llm} introduce a framework showing that many popular models inconsistently refuse unethical persuasion requests and may exhibit manipulative tactics under certain prompts, motivating stronger alignment and policy controls for persuasion-capable systems. Complementing this, benchmark- and platform-oriented efforts such as \citet{singh2024measuring} propose \emph{PersuasionBench} and \emph{PersuasionArena} to systematically evaluate persuasion (including the “transsuasion’’ transformation task), documenting both model strengths in generation and notable weaknesses in simulating persuasion dynamics.

Methodologically, the personalization toolkit around LLMs is expanding. Data- and optimization-efficient approaches include modular parameter sharing as in \citet{tan2024personalized} (PER-PCS), which composes user-specific “pieces’’ for efficient adaptation; prompt-then-edit pipelines as in \citet{zhang2025personalize}, which synthesize self-preference data and modify internal representations for rapid on-the-fly personalization; and preference induction from user writing samples, as in \citet{aroca2025aligning} (PROSE), which yields substantial improvements in stylistic alignment. Benchmarking frameworks such as \citet{zollo2024personalllm} (PersonalLLM) further standardize evaluation of user-tailored outputs under sparse feedback, enabling controlled comparisons across algorithms and data regimes. Broader surveys synthesize these directions: \citet{chen2024large} outline opportunities and challenges for LLM-centric personalized computing (e.g., long-term user modeling, privacy, tool use), while \citet{xu2025personalized} systematize personalized generation across modalities and tasks, identifying open problems in data scarcity, privacy-preserving modeling, and cross-modal alignment.

Applications span beyond persuasion per se to personalized assistance and narrative engagement. In information seeking, \citet{baek2024knowledge} show that knowledge-augmented LLMs can produce more contextually relevant, user-tailored query suggestions; \citet{prahlad2025personalizing} similarly demonstrate GraphRAG-style personalization by integrating user knowledge graphs for dialogue. In narrative domains, \citet{yunusov2024mirrorstories} find that personalized “mirror’’ stories generated by GPT-4 are rated more engaging than generic AI or human baselines, suggesting benefits for education and inclusion. Within synthetic social reasoning, \citet{breum2024persuasive} observe that arguments emphasizing factual evidence, trust, and status cues tend to be most persuasive, offering design signals for LLM-mediated influence.

Advertising is a focal use case and a locus of concern. \citet{meguellati2024good} show that GPT-based models can match human-crafted, persona-tailored ad copy in relevance and appeal, indicating near-term viability of scalable customized creative. Yet user research reveals pitfalls: \citet{tang2024genai} find that embedded, personalized ads in chatbot responses are often unnoticed without disclosure but reduce trust and increase perceived intrusiveness when disclosed, raising UX, transparency, and autonomy questions. Finally, head-to-head tests against experts suggest that state-of-the-art LLMs are already competitive persuaders: \citet{hackenburg2025comparing} report GPT-4 performs on par with political campaign professionals overall (and better on some topics), while persona-alignment yields limited average gains—again tempering expectations about microtargeting’s marginal returns.

\section{METHODOLOGY}
\label{methodology}

\begin{figure}[h]
    \centering
    \includegraphics[width=\linewidth]{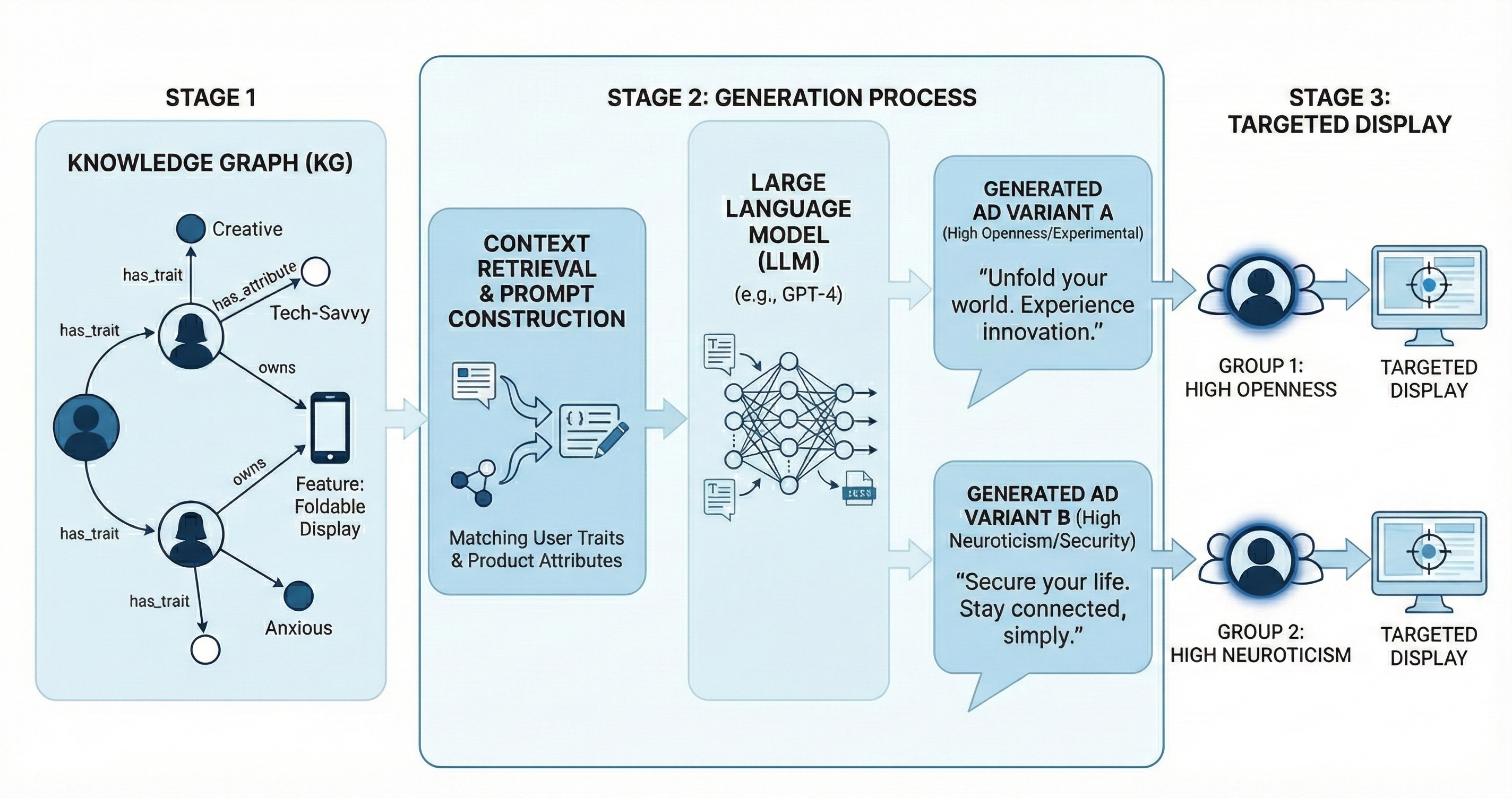}
    \caption{The proposed generation pipeline. Stage 1 utilizes a KG to map user-product relationships, providing structured context for prompt construction. In Stage 2, the LLM transforms these constraints into persuasive narratives (Generative Storytelling), which are then delivered to matched audiences in Stage 3.}
    \label{fig:workflow}
\end{figure}

We employ a generation pipeline that leverages Knowledge Graphs (KGs) to LLMs outputs, as illustrated in Figure \ref{fig:workflow}. Specifically, the KG functions as a structured grounding layer, effectively mapping the semantic relationships between abstract user traits and concrete product attributes to ensure that the prompt context is data-driven rather than generic. In this architecture, the LLM serves as the central engine, transforming structured constraints—from user attributes to psychological principles—into targeted advertising copy. To strictly evaluate the quality and efficacy of these AI-generated narratives without the confounding variables of a live deployment environment, we isolate the generative module in our experiments. This allows us to systematically benchmark the LLM's performance against human experts across both personality-based and persuasion-based tasks.

\subsection{Study 1: Personalizing with LLMs}

This study aims to explore how effective LLM-generated ads are compared to ads created by humans. We are interested in how this effectiveness is tied to personality traits, especially in individuals with high levels of openness or neuroticism. To this end, we focus on: (i)~ studying how well LLMs can personalize ads for specific personality traits, (ii)~ finding out which personality trait works best for LLM-based ad personalization, and (iii)~ comparing how LLM-generated ads and human-written ads perform in terms of personalization and user engagement.

% \subsection{Study Design}\label{sec:design}
% \subsubsection{Survey Overview}
To study the individual preferences of the participants with respect to their ad engagement, we designed two tasks followed by the Big5 questionnaire \citep{condon2022sapa}. 

After showing them an ad,
\textit{Task 1} is designed to capture participants' responses on a 5-point Likert scale for three aspects: (i)~the product attitude, where we ask them to provide a rating to indicate how they like this advertisement, (ii)~the purchase intention, where we ask them the likelihood of their intent to purchase the advertised product after viewing the advertisement, and (iii)~the engagement intention, assessing their interest in seeking further information about the product featured in the advertisement. 
\textit{Task 2} seeks to assess the effectiveness of LLM-generated ads compared to human-written ads when presented side by side. It aims to examine participant preferences in selecting the best ads. By displaying both types of ads simultaneously, we can evaluate their relative performance and determine how generated ads comparatively appeal to human-written ads. 
\textit{The Big5} is a 20-item questionnaire that allows us to calculate participants' personality scores and it is administered after subjects complete the previous tasks. 

\subsubsection{\textbf{Dataset}}

The human-written ads used in our study were sourced from the research  by \citet{winter2021effects}. For the generation of ad messages targeting openness and neuroticism traits, we employed two LLM prompts. The prompts are as follows:

\begin{itemize}
  \item ``write a 1 line ad for a phone called Xphone targeting people with the openness trait without mentioning the trait explicitly"
  
  \item ``write a 1 line ad for a phone called Xphone targeting people with the neuroticism trait without mentioning the trait explicitly"
\end{itemize}

In our experiments, we maintained the `temperature' – a setting of  that controls the randomness of the model's response - at its default setting of 0.7. This value offers a balance between randomness and determinism which provides the model with enough flexibility to generate inventive outputs, while still ensuring it stays relevant to the given prompts.

As the LLM, we utilized GPT 3.5, despite the availability of newer models such as GPT 3.5-Turbo (also known as ChatGPT) and GPT-4. Our decision was motivated by the desire for reproducibility and consistency as during our study both GPT 3.5-Turbo and GPT-4 were undergoing frequent updates.

\begin{figure}[t]
    \centering
    \includegraphics[width=\columnwidth]{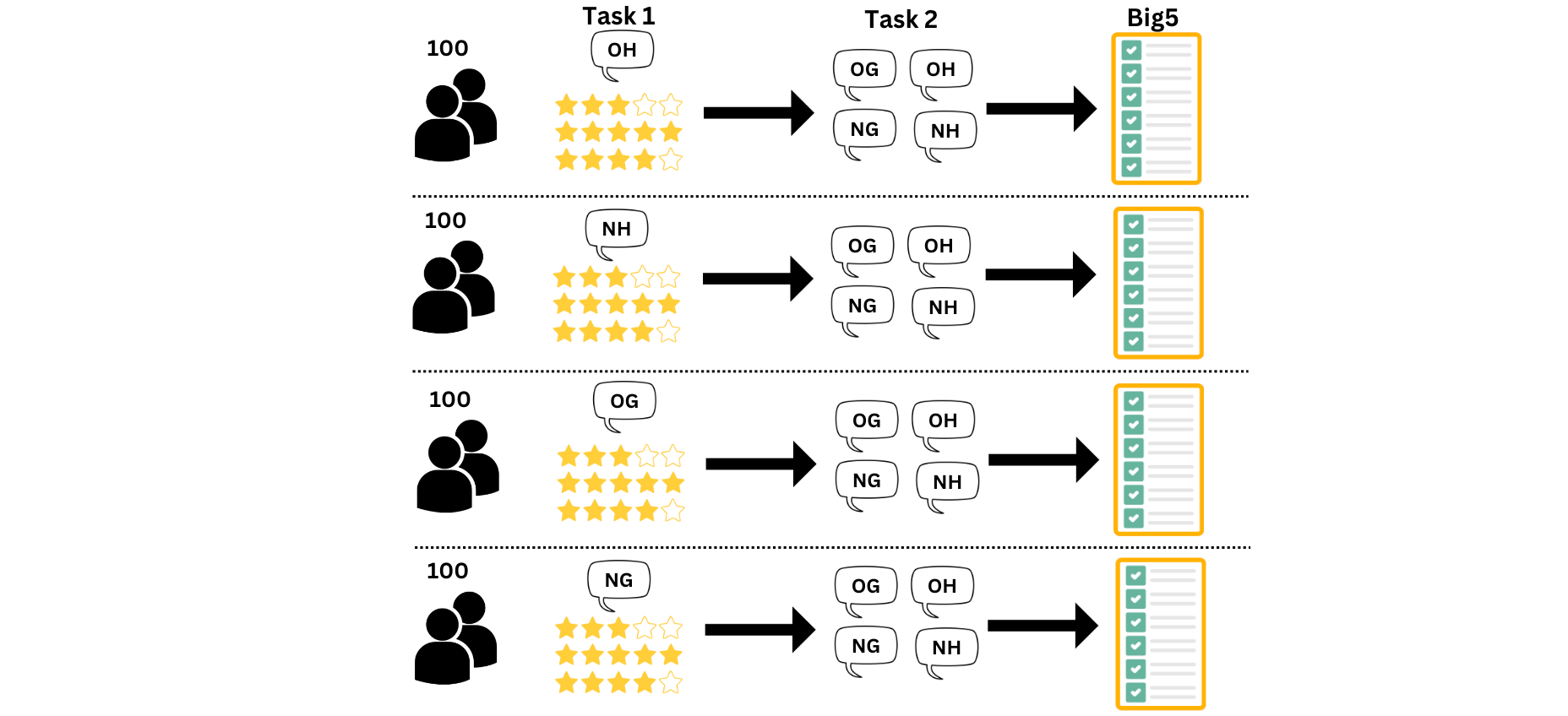} % Adjust the width to fit within a single column
    \caption{Procedure of the experiment: (i) Task 1, including three questions presented on a 5-point Likert scale; (ii)~Task 2, prompting participants to select one ad message from the four messages displayed side-by-side; followed by (iii)~the Big5 questionnaire to gather their personality information. Note that there are four variants of Task 1: OH, OG, NH, and NG}
    \label{fig:survey_overview}
\end{figure}

\subsubsection{\textbf{Task Design}}
\paragraph{Task 1}
Four surveys were created as shown in Figure \ref{fig:survey_overview}. The study included multiple versions of the same ad image with varying messages customized for individuals with high openness or neuroticism traits, and created either by a human, or the LLM.

% Table \ref{table:your_label}.

Participants were presented with the same image in all surveys along with one of the four message variations, where the survey type is associated with the ad message shown  as follows:
    
        \begin{itemize}
        \item \textbf{OH}: \textbf{O}penness, written by a \textbf{H}uman 
        \item \textbf{OG}: \textbf{O}penness, \textbf{G}enerated by an LLM
        \item \textbf{NH}: \textbf{N}euroticism, written by a \textbf{H}uman
        \item \textbf{NG}: \textbf{N}euroticism, \textbf{G}enerated by an LLM
        \end{itemize}

    Participants were then asked to rate their perception of the product, purchase intention, and engagement intention on a 5-point Likert scale ranging from ``very unlikely" to ``very likely".
    
\paragraph{Task 2} Once they completed Task 1, participants were shown all four messages simultaneously and asked to select their preferred message. It is important to consider the potential bias that may arise from participants having seen one of the four ads in Task 1. However, this bias is balanced out across the four surveys since an equal number of participants were exposed to each ad before proceeding to Task 2. By ensuring that the same number of participants viewed each of the four ads before the preference selection task, the influence of any individual ad exposure in Task 1 is mitigated. Thus, when examining the data in an aggregate fashion, the bias resulting from the initial exposure to a specific ad is minimized, ensuring a fair assessment of participants' preferences in Task 2.

\paragraph{Big5 questionnaire} After having completed the tasks, participants answered a Big Five personality questionnaire (20 items) focusing on openness and neuroticism \citep{condon2022sapa}. The analysis explored the relationship between these traits and task responses. Participants were categorized based on their openness and neuroticism scores. The questionnaire was strategically placed at the study's end to minimize bias from prior exposure to the questionnaire, though it leads to reduced control over personality and demographic distributions.

\subsubsection{\textbf{Measurements}}
\paragraph{Task 1} The Mann-Whitney test was utilized to assess statistical significance because the data is not equally distributed, and the Benjamini-Hochberg correction was applied to account for multiple comparisons.
\paragraph{Task 2} We employed the chi-square test to examine variable independence and assess the relationship between the different ad messages.
\paragraph{Personality Traits} To systematically compare the effectiveness of generated ads with matched and non-matched personality traits, this study measures the Big Five personality traits used to tailor messages. These traits are measured with The Big Five personality questionnaire \citep{goldberg1999broad, condon2022sapa} on a 6-point Likert scale, including openness (e.g., ``Am able to come up with new and different ideas") and neuroticism (e.g., ``Get overwhelmed by emotions"), as they are the focus of the study.

\subsubsection{\textbf{Participants}}

The study recruited 400 participants from Amazon Mechanical Turk (MTurk), with an equal distribution of 100 participants assigned to each survey variation. The participants' demographics are: 297 males (74.2\%) and 103 females (25.8\%), with an average age of 32. The tasks took approximately 5 minutes to complete, and participants were rewarded \$1 for their time. To ensure English proficiency, participants must be 18 or older and reside in the United States. 

\subsection{Study 2: Persuasion Principles with LLMs}

This study investigates the effectiveness of AI-generated advertisements compared to human-created advertisements across different persuasion principles \citep{Cialdini2004}. While Study 1 focused on personality-based personalization with varying text messages, Study 2 examines complete advertisements (both image and text) designed around established persuasion principles. Our objectives are to: (i) assess participant preferences between AI-generated and human-created ads within different persuasion contexts, (ii) understand the qualitative factors influencing these preferences, and (iii) evaluate participants' ability to distinguish AI-generated content from human-created content.

\begin{figure}[t]
    \centering
    \includegraphics[width=\columnwidth]{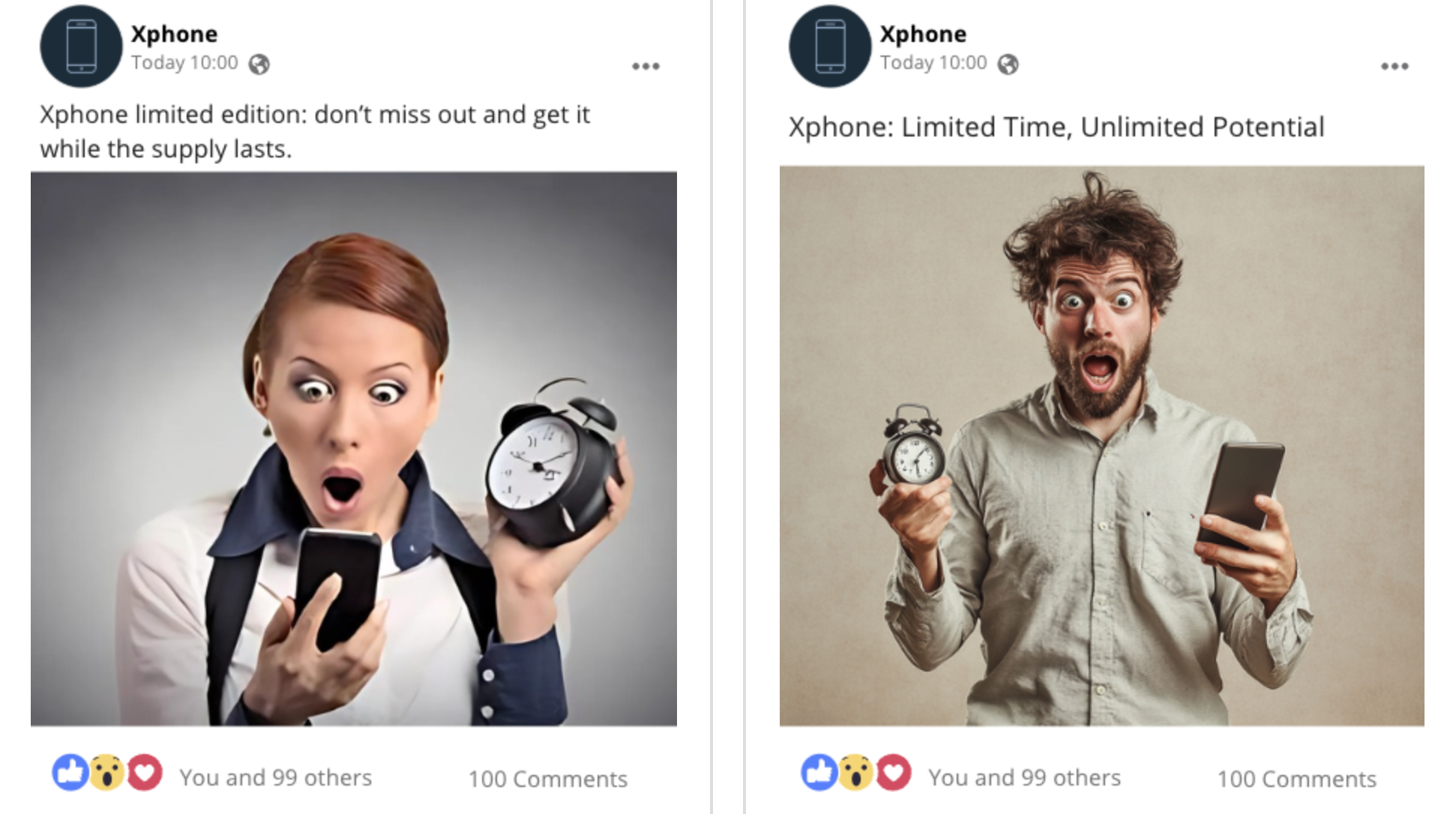} % Adjust the width to fit within a single column
    \caption{Example of advertisements used in Study 2 for the scarcity persuasion principle condition. Left: Human-created advertisement. Right: AI-generated advertisement (text generated using LLMs, image generated using a diffusion model - Midjourney). Both ads emphasize time-limited availability and urgency, core elements of the scarcity principle. Participants viewed both advertisements side-by-side without knowledge of which was AI-generated.}
    \label{fig:s2_overview}
\end{figure}

\subsubsection{\textbf{Dataset}}

For each persuasion condition, we generated outputs from three LLMs (GPT-5, Sonnet 4, LLama 4). To rigorously test the democratization potential of GenAI, we adopted a \textit{non-expert selection protocol}. A single author (with a Computer Science background and no professional advertising training) manually selected the final variant based on subjective appeal and perceived quality, acting as a proxy for a general end-user.
This methodological choice ensures our results represent a \textit{conservative lower bound} of AI performance: if a non-expert selector relying on intuition can achieve parity or superiority, professional curation would likely yield even higher efficacy.

All advertisements featured the same product (Xphone) used throughout Study 1, ensuring consistency across both studies. The advertisements were designed around four distinct persuasion principles \citep{Cialdini2004}:

\begin{itemize}
  \item \textbf{Consensus}: Emphasizing social proof and the popularity of the product among users
  \item \textbf{Authority}: Highlighting expert endorsements and professional credibility
  \item \textbf{Scarcity}: Focusing on limited availability and time-sensitive offers
  \item \textbf{Cognition}: Appealing to rational thinking and logical product features
\end{itemize}

\subsubsection{\textbf{Task Design}}

We employed a between-subjects design with four experimental conditions corresponding to the four persuasion principles. A total of 800 participants were recruited, with 200 participants randomly assigned to each condition. Unlike Study 1, where participants evaluated advertisements with identical images but varying text messages, Study 2 presented participants with complete advertisements featuring both unique images and text tailored to each persuasion principle. Figure~\ref{fig:s2_overview} illustrates an example of the advertisements used in the scarcity condition, showing both the human-created and AI-generated versions presented to participants.

In each condition, participants were shown two advertisements side-by-side: one human-created and one AI-generated. Both advertisements targeted the same persuasion principle and featured the Xphone product. The presentation order of the human-created and AI-generated ads was randomized to control for position bias.

Participants completed four sequential tasks:

\paragraph{Question 1: Preference Selection} Participants were asked to select which of the two advertisements they preferred. This forced-choice format allowed us to measure direct preference between AI-generated and human-created content within each persuasion context.

\paragraph{Question 2: Preference Reasoning} Following their selection, participants provided an open-ended explanation for their preference. This qualitative data enabled us to identify the underlying reasons driving participant choices and to understand which advertisement characteristics resonated most strongly with viewers.

\paragraph{Question 3: Influencing Factors} Participants were then asked to describe, in an open-ended format, the specific factors that influenced their decision. This question aimed to elicit more detailed insights into the design elements, messaging strategies, or other attributes that contributed to their preference formation.

\paragraph{Question 4: AI Detection} Finally, participants were asked to identify which of the two advertisements they believed was generated by AI. It is important to note that participants were not informed about the AI-generated nature of any advertisement during Questions 1-3, ensuring their preference selection and reasoning were unbiased by knowledge of the ad's origin. This design allows for subsequent analysis examining potential bias towards or against AI-generated content by comparing participants' actual preferences (Question 1) with their perceptions of which ad was AI-generated (Question 4). Such analysis can reveal whether participants exhibit systematic preference patterns based on perceived AI involvement, independent of actual ad quality or effectiveness.

\subsubsection{\textbf{Measurements}}
\paragraph{Advertisement Preference} The chi-square test ($\chi^2$) was utilized to assess the relationship between advertisement type (AI vs. Human) and participant preference, as well as to test independence across conditions.  Bonferroni correction was applied to account for multiple comparisons across the four persuasion conditions (adjusted significance threshold: $\alpha = 0.0125$, calculated as 0.05/4).

\paragraph{Effect Sizes} Cohen's h was calculated for overall proportional differences between AI and human preference rates. Cohen's d was computed for condition-specific effect sizes (Authority, Consensus, Cognition, Scarcity). Risk Ratios (RR) with 95\% confidence intervals were used to quantify the relative likelihood of preferring AI versus human advertisements.

\paragraph{Detection Accuracy} The Kappa statistic ($\kappa$) was employed to measure agreement between participants' AI identification and actual advertisement origin beyond chance. Odds Ratios (OR) with 95\% confidence intervals were calculated to assess the impact of AI detection on preference rates.

\paragraph{Qualitative Analysis} To analyze the open-ended justifications (Questions 2 and 3), we employed an inductive thematic analysis approach. Responses were coded to identify recurrent drivers of preference, such as message sophistication and visual coherence. A detailed breakdown of the thematic definitions, coding methodology, and supplementary participant quotes is provided in Appendix~\ref{sec:appendix_qualitative}.

\subsubsection{\textbf{Participants}}
The study recruited 800 participants from Prolific, with an equal distribution of 200 participants assigned to each of the four persuasion principle conditions (Authority, Consensus, Cognition, Scarcity). The participant demographics were: 362 females (45.3\%) and 432 males (54.0\%), with ages ranging from 18 to 65+. The tasks took approximately 2 minutes to complete, and participants were compensated £0.50 for their time (equivalent to £15/hour). To ensure data quality, participants were required to be 18 or older, have English as their first language, and reside in the United States.

\paragraph{\textbf{Ethical considerations}} Both studies were conducted anonymously, and participant data is kept confidential. The studies have been approved by the first author's institution IRB. Participants provided informed consent to collect and analyze their data. Compensation rates for both studies exceeded recommended minimum wages.

\section{RESULTS}
\label{sec:results}

Our evaluation strategy mirrors the paper's progression from investigating data-driven personalization to exploring universal persuasion.

In \textbf{Study 1}, the objective is to determine if GenAI can effectively tailor content based on explicit user data (Big Five traits). We utilize psychometric Likert scales (\textit{Product Rating, Purchase Intention}) to measure the \textit{depth of resonance}—verifying that the model is not just producing generic text, but successfully aligning with the specific psychological profile of the user.

In \textbf{Study 2}, the focus shifts to investigating the model's capacity to apply universal persuasion principles and generate multimodal assets (text and visuals) \textit{without} prior knowledge of the user. Here, we transition to a forced-choice metric (\textit{Preference Rate}) to evaluate the \textit{comparative efficacy} of AI against human experts in a competitive setting.

While this difference in measurement precludes a direct statistical comparison of effect sizes between the two studies, it allows us to evaluate each capability within its appropriate context: validating the \textit{internal fit} of trait-based text in the former, and assessing the \textit{external impact} of multimodal persuasion in the latter.

\subsection{Study 1: Personalizing with LLMs}
\label{sec:results_study1}

\subsubsection{Task 1: Effectiveness of Personality-Tailored Ads}

In the first task, we evaluated the efficacy of advertisement messages generated by LLMs compared to those written by humans, specifically examining the impact of personality congruence. We measured three key dimensions of user engagement: \textit{Product Rating}, \textit{Purchase Intention}, and \textit{Engagement Intention}. 

To ensure robust statistical analysis, we employed the Mann-Whitney U test to assess differences between groups, given the non-normal distribution of the Likert-scale data. To control for the False Discovery Rate (FDR) accumulation due to multiple hypothesis testing, we applied the Benjamini-Hochberg correction to all reported $p$-values.

\begin{table}[ht]
\centering
\caption{Comparative analysis of ad effectiveness (Means) and statistical significance (P-values) for Matched vs. Unmatched personality traits. Significant values ($p \le 0.05$) are highlighted in bold.}
\label{tab:study1_combined}
\small
\setlength{\tabcolsep}{4pt}
\renewcommand{\arraystretch}{1.3}
\begin{tabular}{@{}llccc@{}}
\toprule
\textbf{Condition} & \textbf{Measurement} & \textbf{Matched} & \textbf{Unmatched} & \textbf{\textit{p}} \\
\midrule
\multicolumn{5}{@{}l@{}}{\textit{\textbf{Generated Ads (LLM)}}} \\
\midrule
\multirow{3}{*}{\shortstack[l]{\textbf{Openness}\\\textbf{(OG)}}} 
  & Product Rating & 4.14 & 3.71 & \textbf{0.02} \\
  & Purchase Intention & 4.14 & 3.69 & \textbf{0.02} \\
  & Engagement Intention & 4.33 & 3.73 & \textbf{0.01} \\
\addlinespace[0.3em]
\multirow{3}{*}{\shortstack[l]{\textbf{Neuroticism}\\\textbf{(NG)}}} 
  & Product Rating & 3.84 & 4.00 & 0.33 \\
  & Purchase Intention & 3.77 & 4.15 & 0.27 \\
  & Engagement Intention & 3.97 & 4.29 & 0.27 \\
\midrule
\multicolumn{5}{@{}l@{}}{\textit{\textbf{Human-Written Ads (Baseline)}}} \\
\midrule
\multirow{3}{*}{\shortstack[l]{\textbf{Openness}\\\textbf{(OH)}}} 
  & Product Rating & 4.13 & 3.96 & 0.50 \\
  & Purchase Intention & 4.33 & 3.68 & \textbf{0.05} \\
  & Engagement Intention & 4.30 & 3.88 & 0.15 \\
\addlinespace[0.3em]
\multirow{3}{*}{\shortstack[l]{\textbf{Neuroticism}\\\textbf{(NH)}}} 
  & Product Rating & 3.61 & 3.76 & 0.54 \\
  & Purchase Intention & 3.74 & 4.00 & 0.54 \\
  & Engagement Intention & 3.71 & 4.15 & 0.47 \\
\bottomrule
\end{tabular}
\end{table}

\textbf{Impact of Personality Congruence (Matched vs. Unmatched)}\\
Table~\ref{tab:study1_combined} presents a consolidated view of the mean scores and statistical significance for both LLM-generated and human-written advertisements.

For the \textbf{Openness} trait, we observed a strong positive effect of personalization. In the LLM-generated condition (OG), participants with matching personality traits reported significantly higher scores across all three metrics compared to unmatched participants. Specifically, Product Rating ($p=0.02$), Purchase Intention ($p=0.02$), and Engagement Intention ($p=0.01$) all showed statistically significant improvements when the ad content was congruent with the user's Openness level. A similar trend was observed in the human-written control condition (OH), where Purchase Intention was significantly higher for matched participants ($p=0.05$), reinforcing that the Openness trait is highly receptive to personalized messaging regardless of the authoring source.

Conversely, the results for the \textbf{Neuroticism} trait revealed a more complex dynamic. As detailed in Table~\ref{tab:study1_combined}, matching ad content to high-neuroticism users did not yield a positive increase in engagement. In several instances, such as the Generated Neuroticism (NG) condition, matched participants actually reported lower Purchase Intentions ($M=3.77$) compared to unmatched participants ($M=4.15$), though these differences did not reach statistical significance after correction ($p > 0.05$). This finding aligns with prior literature suggesting that individuals with high neuroticism scores may exhibit avoidance behaviors or heightened caution when targeted with stimuli that resonate with their trait-specific anxieties. Consequently, while LLMs successfully generated content that aligned with the trait theoretically, the \textit{receptiveness} of the target audience to neuroticism-based appeals remains low.

\textbf{Comparative Efficacy: AI vs. Human}\\
A critical objective of this study was to determine if LLM-generated content could rival human creativity. We conducted a direct comparison between the LLM-generated ads (OG/NG) and human-written ads (OH/NH) specifically for the matched personality groups. The statistical analysis revealed no significant differences across any of the three measured dimensions ($p > 0.05$ for all comparisons). For example, in the Openness condition, the Product Rating for LLM ads ($M=4.14$) was statistically indistinguishable from human ads ($M=4.13$). This lack of significant divergence suggests that basic LLM prompting strategies (zero-shot) are sufficient to produce advertising copy that is perceptually equivalent to content crafted by human experts in terms of user engagement and purchase intent.

\subsubsection{Task 2: Preference and Competitive Performance}

In the second task, participants evaluated the four ad variations simultaneously in a simulated competitive shopping environment. This forced-choice design allowed us to assess relative preference and potential click-through behavior.

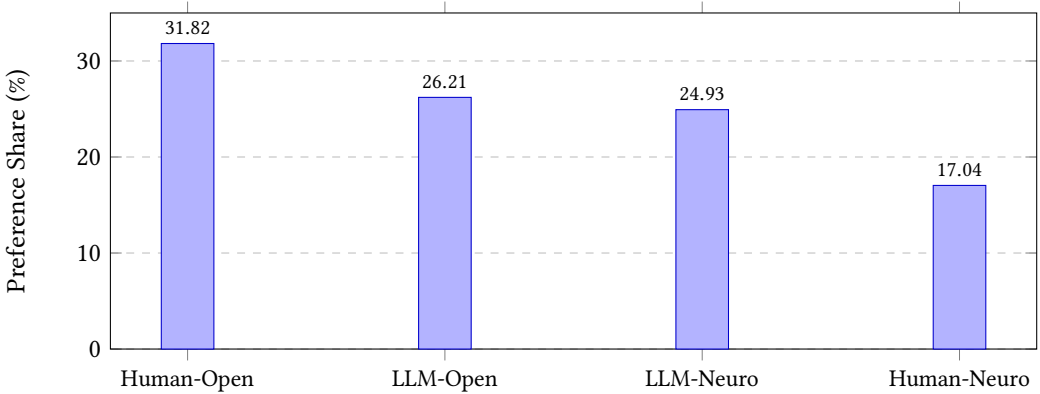
\begin{figure}[ht]
\centering
\begin{tikzpicture}
\begin{axis}[
    ybar,
    width=\columnwidth,
    height=6cm,
    bar width=0.7cm,
    ylabel={Preference Share (\%)},
    symbolic x coords={Human-Open, LLM-Open, LLM-Neuro, Human-Neuro},
    xtick=data,
    x tick label style={font=\small},
    ymin=0,
    ymax=35,
    nodes near coords,
    nodes near coords align={vertical},
    every node near coord/.append style={font=\footnotesize},
    enlarge x limits=0.10,
    ymajorgrids=true,
    grid style=dashed,
]
\addplot[fill=blue!30, draw=blue!80!black] coordinates {
    (Human-Open, 31.82)
    (LLM-Open, 26.21)
    (LLM-Neuro, 24.93)
    (Human-Neuro, 17.04)
};
\end{axis}
\end{tikzpicture}
\caption{Distribution of participant preferences (clicks) when ads were presented side-by-side in Task 2. LLM-generated ads collectively received 51.14\% of clicks compared to 48.86\% for human-written ads.}
\label{fig:study1_clicks}
\end{figure}

The distribution of user clicks is visualized in Figure~\ref{fig:study1_clicks}. The human-written ad tailored to Openness emerged as the most preferred option, garnering 31.82\% of total clicks. However, the LLM-generated ad for Openness followed closely with 26.21\%. Notably, the LLM-generated ad for Neuroticism (24.93\%) significantly outperformed its human-written counterpart (17.04\%), suggesting that the LLM may have captured a nuance in the neuroticism prompt that resonated more effectively---or was less aversive---than the human attempt in a comparative setting.

A Chi-Squared test for independence confirmed a significant association between ad type and user preference ($\chi^2$, $p < 0.05$). When aggregating the data by source, LLM-generated ads secured a total of 51.14\% of clicks, marginally surpassing human-written ads at 48.86\%. These results indicate that LLM-generated content is not only comparable in isolation (Task 1) but remains competitive when presented alongside human-authored alternatives, demonstrating high stability across different trait targets.

\subsection{Study 2: Persuasion Principles with LLMs}
\label{sec:results_study2}

\subsubsection{Advertisement Preference and Effectiveness}

In the second study, we evaluated the comparative effectiveness of AI-generated versus human-created advertisements across four distinct persuasion principles. 

\paragraph{Overall Performance}
AI-generated advertisements achieved a dominant 59.1\% preference rate compared to 40.9\% for human-created advertisements ($\chi^2 = 26.65$, $p < 0.001$). The effect size was medium (Cohen's $h = 0.37$), representing a substantial 18.2 percentage point advantage for the AI-generated content.

\paragraph{Performance by Persuasion Principle}
As detailed in Table~\ref{tab:study2_performance}, performance varied significantly depending on the persuasion strategy employed. After applying the Bonferroni correction for multiple comparisons ($\alpha = 0.0125$), AI advertisements maintained a statistically significant advantage in three of the four conditions. The strongest performance was observed in the \textit{Authority} ($d = 0.52$) and \textit{Consensus} ($d = 0.50$) conditions, suggesting that LLMs are particularly adept at generating content that leverages credibility and social proof. Conversely, the \textit{Scarcity} condition showed no meaningful difference ($p = 0.671$), with qualitative feedback indicating a general consumer skepticism toward urgency tactics regardless of the source.

\begin{table}[ht]
\centering
\caption{AI advertisement performance across persuasion principles. Effect sizes (Cohen's \textit{d}) and Risk Ratios (RR) are reported. Significant values after Bonferroni correction ($p \le 0.0125$) are bolded.}
\label{tab:study2_performance}
\small
\setlength{\tabcolsep}{4pt}
\renewcommand{\arraystretch}{1.2}
\begin{tabular}{@{}lcccccc@{}}
\toprule
\textbf{Principle} & \textbf{Human} & \textbf{AI} & \textbf{Gap} & \textbf{\textit{p}-value} & \textbf{Cohen's \textit{d}} & \textbf{RR [95\% CI]} \\
\midrule
Authority & 37.0\% & 63.0\% & +26.0\% & \textbf{<0.001} & 0.52 & 1.70 [1.32--2.19] \\
Consensus & 37.5\% & 62.5\% & +25.0\% & \textbf{<0.001} & 0.50 & 1.67 [1.30--2.14] \\
Cognition & 40.5\% & 59.5\% & +19.0\% & \textbf{0.007} & 0.38 & 1.47 [1.15--1.87] \\
Scarcity & 48.5\% & 51.5\% & +3.0\% & 0.671 & 0.06 & 1.06 [0.86--1.31] \\
\bottomrule
\end{tabular}
\end{table}

\paragraph{Demographic Homogeneity in Preference}
Table~\ref{tab:study2_demographics} presents the demographic breakdown of advertisement preferences. Statistical analysis revealed no significant effects of age ($p = 0.363$) or gender ($p = 0.827$) on advertisement choice. This indicates that the preference for AI-generated aesthetics and messaging was consistent across demographic cohorts, suggesting a broad appeal rather than a niche trend among specific user groups.

\begin{table}[ht]
\centering
\caption{Demographic distribution of advertisement preferences. AI preference rates (``AI Choosers'') remained stable across age groups and gender categories.}
\label{tab:study2_demographics}
\small
\setlength{\tabcolsep}{8pt}
\renewcommand{\arraystretch}{1.2}
\begin{tabular}{@{}lcccc@{}}
\toprule
\textbf{Demographic} & \textbf{\textit{N}} & \textbf{Chose AI} & \textbf{Chose Human} & \textbf{AI Preference \%} \\
\midrule
\multicolumn{5}{@{}l@{}}{\textit{\textbf{Age Group}}} \\
\midrule
18--24 & 32 & 20 & 12 & 62.5\% \\
25--34 & 168 & 102 & 66 & 60.7\% \\
35--44 & 198 & 117 & 81 & 59.1\% \\
45--54 & 190 & 109 & 81 & 57.4\% \\
55--64 & 136 & 77 & 59 & 56.6\% \\
65+ & 76 & 48 & 28 & 63.2\% \\
\midrule
\multicolumn{5}{@{}l@{}}{\textit{\textbf{Gender}}} \\
\midrule
Female & 362 & 218 & 144 & 60.2\% \\
Male & 432 & 251 & 181 & 58.1\% \\
\midrule
\textbf{Total Sample} & \textbf{800} & \textbf{473} & \textbf{327} & \textbf{59.1\%} \\
\bottomrule
\end{tabular}
\end{table}

\paragraph{Qualitative Drivers of Preference}
Analysis of open-ended responses identified three primary drivers behind the success of AI advertisements:
\begin{itemize}
    \item \textbf{Message Sophistication:} AI advertisements frequently employed aspirational language. Participants described these messages as ``inspiring'' (36 mentions) and ``professional'' (44 mentions).
    \item \textbf{Visual Coherence:} 32\% of qualitative responses noted superior color palettes and image-text unity in the AI ads. One participant notably described an AI ad's aesthetic as evoking ``a professor's lecture room,'' reinforcing the Authority principle.
    \item \textbf{Psychological Precision:} AI demonstrated a superior application of persuasion principles, particularly in Authority and Consensus conditions, where participants valued clear markers of ``credentials'' and ``diverse group representation.''
\end{itemize}

\subsubsection{AI Detection Capabilities}

We assessed participants' ability to distinguish between human and AI-generated content (``AI Spotters''). Overall, 58.4\% of participants correctly identified the AI-generated advertisement (Kappa $= 0.17$), indicating only a moderate detection ability beyond random chance.

However, as shown in Table~\ref{tab:study2_detection}, detection accuracy was heavily stratified by age. Younger participants (18--24) demonstrated significantly higher detection rates (72.5\%) compared to the oldest cohort (65+), who performed near chance levels (49.2\%). The odds of correctly identifying AI were 2.78 times higher for the youngest group compared to the oldest (OR $= 2.78$, 95\% CI: [1.23, 6.28], $p < 0.01$).

\begin{table}[ht]
\centering
\caption{AI detection accuracy (``AI Spotters'') broken down by persuasion principle and age group.}
\label{tab:study2_detection}
\small
\setlength{\tabcolsep}{6pt}
\renewcommand{\arraystretch}{1.2}
\begin{tabular}{@{}lcccc@{}}
\toprule
\textbf{Category} & \textbf{\textit{N}} & \textbf{Correct ID} & \textbf{Incorrect ID} & \textbf{Accuracy \%} \\
\midrule
\multicolumn{5}{@{}l@{}}{\textit{\textbf{By Persuasion Principle}}} \\
\midrule
Consensus & 200 & 151 & 49 & 75.5\% \\
Cognition & 200 & 129 & 71 & 64.5\% \\
Authority & 200 & 125 & 75 & 62.5\% \\
Scarcity & 200 & 62 & 138 & 31.0\% \\
\midrule
\multicolumn{5}{@{}l@{}}{\textit{\textbf{By Age Group}}} \\
\midrule
18--24 & 32 & 23 & 9 & 72.5\% \\
25--34 & 168 & 113 & 55 & 67.3\% \\
35--44 & 198 & 124 & 74 & 62.8\% \\
45--54 & 190 & 113 & 77 & 59.3\% \\
55--64 & 136 & 74 & 62 & 54.6\% \\
65+ & 76 & 37 & 39 & 49.2\% \\
\midrule
\textbf{Overall} & \textbf{800} & \textbf{467} & \textbf{333} & \textbf{58.4\%} \\
\bottomrule
\end{tabular}
\end{table}

\subsubsection{The Perception-Choice Interaction}

To understand whether identifying an advertisement as AI-generated influenced user preference, we analyzed the interaction between perception and choice. Table~\ref{tab:study2_perception_matrix} presents a matrix of these behaviors.

\paragraph{The Bias Penalty vs. Quality Resilience}
To quantify the bias, we calculated the \textit{Detection Penalty} ($\Delta_{bias}$) as the difference in AI preference rates between participants who were deceived and those who correctly identified the source:

\begin{equation}
    \Delta_{bias} = P(\text{Choose AI} | \text{Believed Human}) - P(\text{Choose AI} | \text{Detected AI})
\end{equation}

Substituting the observed rates yields a significant penalty:

$$ \Delta_{bias}  21.2\% = 71.5\% - 50.3\%   $$

This calculation confirms a significant bias against known AI content ($OR = 0.40$, $p < 0.001$). However, critically, even after this penalty is applied, the AI preference rate ($50.3\%$) remains above the actual human advertisement performance ($40.9\%$), demonstrating that execution quality effectively overrides the negative origin bias.

\begin{table}[ht]
\centering
\caption{Choice-Perception Matrix: The relationship between advertisement preference and origin identification. Percentages indicate the proportion of that choice group (e.g., 49.7\% of AI Choosers knew it was AI).}
\label{tab:study2_perception_matrix}
\small
\setlength{\tabcolsep}{8pt}
\renewcommand{\arraystretch}{1.3}
\begin{tabular}{@{}llccc@{}}
\toprule
\textbf{Choice Group} & \textbf{Perception State} & \textbf{\textit{N}} & \textbf{\% of Total} & \textbf{\% of Choice} \\
\midrule
\multirow{2}{*}{\textbf{AI Choosers}} 
  & Correctly identified as AI & 235 & 29.4\% & 49.7\% \\
  & Thought it was human & 238 & 29.8\% & 50.3\% \\
\cmidrule{2-5}
  & \textit{Subtotal} & \textit{473} & \textit{59.1\%} & \textit{100.0\%} \\
\midrule
\multirow{2}{*}{\textbf{Human Choosers}} 
  & Correctly identified as human & 232 & 29.0\% & 71.0\% \\
  & Thought it was AI & 95 & 11.9\% & 29.0\% \\
\cmidrule{2-5}
  & \textit{Subtotal} & \textit{327} & \textit{40.9\%} & \textit{100.0\%} \\
\midrule
% \textbf{Total} & & \textbf{800} & \textbf{100.0\%} & \\
% \bottomrule
\end{tabular}
\end{table}

\paragraph{Synthesizing Quantitative and Qualitative Insights}
Critically, 41.3\% of participants (330 people) made origin-independent choices: either selecting AI while correctly identifying it (29.4\%) or selecting human content while believing it was AI (11.9\%). In both cases, the participant willingly chose an advertisement they \textit{perceived} to be AI-generated. This suggests that for a substantial segment of users, execution quality transcends origin concerns.

Qualitative analysis clarified the demographic divergence observed in the quantitative detection data (Table~\ref{tab:study2_detection}). Younger participants, who were more accurate at spotting AI, tended to use language emphasizing ``vibe'' and ``aesthetic'' with less concern for authenticity. In contrast, older participants placed higher emphasis on ``authenticity'' and ``natural qualities.'' Paradoxically, this search for authenticity often led older participants to misidentify high-quality AI content as human (the "Thought it was human" group), assuming that the sophisticated execution must have been human-made.

\section{DISCUSSION}
\label{sec:discussion}

Our investigation across two studies provides a comprehensive view of Generative AI's role in digital advertising. From text-based personalization (Study 1) to multimodal persuasion (Study 2), the results challenge the traditional assumption that human creativity is the gold standard for marketing communication. Below, we discuss the implications of these findings through five key questions.

\subsection{Can LLMs Effectively Personalize at Scale?}
The results from Study 1 suggest that the answer is a nuanced ``yes,'' but it depends heavily on the target trait. We observed that LLMs excelled at targeting the \textbf{Openness} trait. Participants high in openness—who are naturally more receptive to novelty and innovation—responded significantly better to AI-generated messages than generic ones. In this context, the LLM successfully mimicked the creative flexibility required to engage this audience.
However, the failure to generate effective ads for the \textbf{Neuroticism} trait highlights a fundamental psychological barrier rather than just a model limitation. Despite the LLM technically following the prompt, the resulting ads did not increase engagement for high-neuroticism individuals. This aligns with \textbf{Regulatory Focus Theory} \cite{higgins1997regulatory}, which distinguishes between ``promotion'' and ``prevention'' orientations. While Openness is a \textit{promotion-focused} trait (seeking gain), Neuroticism is a \textit{prevention-focused} trait (avoiding risk). Consequently, simply mirroring the trait in ad copy—without explicitly signaling safety—likely triggers defensive avoidance behaviors regardless of whether the author is human or machine. Thus, while AI enables personalization at scale, it is not yet a universal key for all psychological profiles.

\begin{figure}[H]
    \centering
    % Left Plot: Qualitative Themes (Wider - 60%)
    \begin{minipage}[b]{0.60\linewidth}
        \centering
        % Using the filename for the uploaded bar chart
        \includegraphics[width=\linewidth]{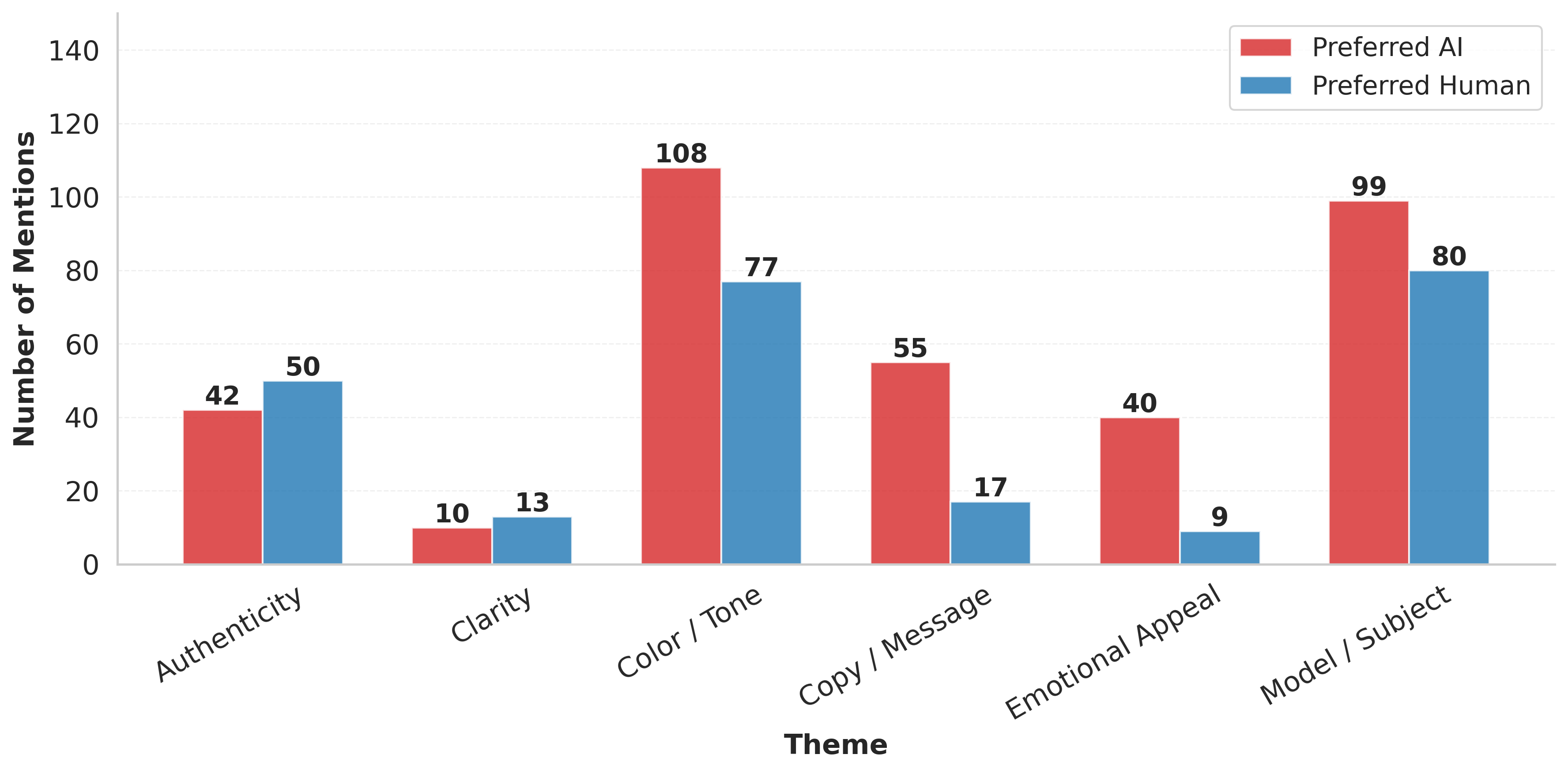} 
        \caption{Qualitative drivers of preference. Participants preferring AI (red) overwhelmingly cited \textbf{Color/Tone} and \textbf{Model Quality}, whereas those preferring humans (blue) prioritized \textbf{Authenticity} and \textbf{Clarity}.}
        \label{fig:qualitative_themes}
    \end{minipage}
    \hfill % Adds spacing between the two plots
    % Right Plot: Dumbbell (Narrower - 38%)
    \begin{minipage}[b]{0.38\linewidth}
        \centering
        \includegraphics[width=\linewidth]{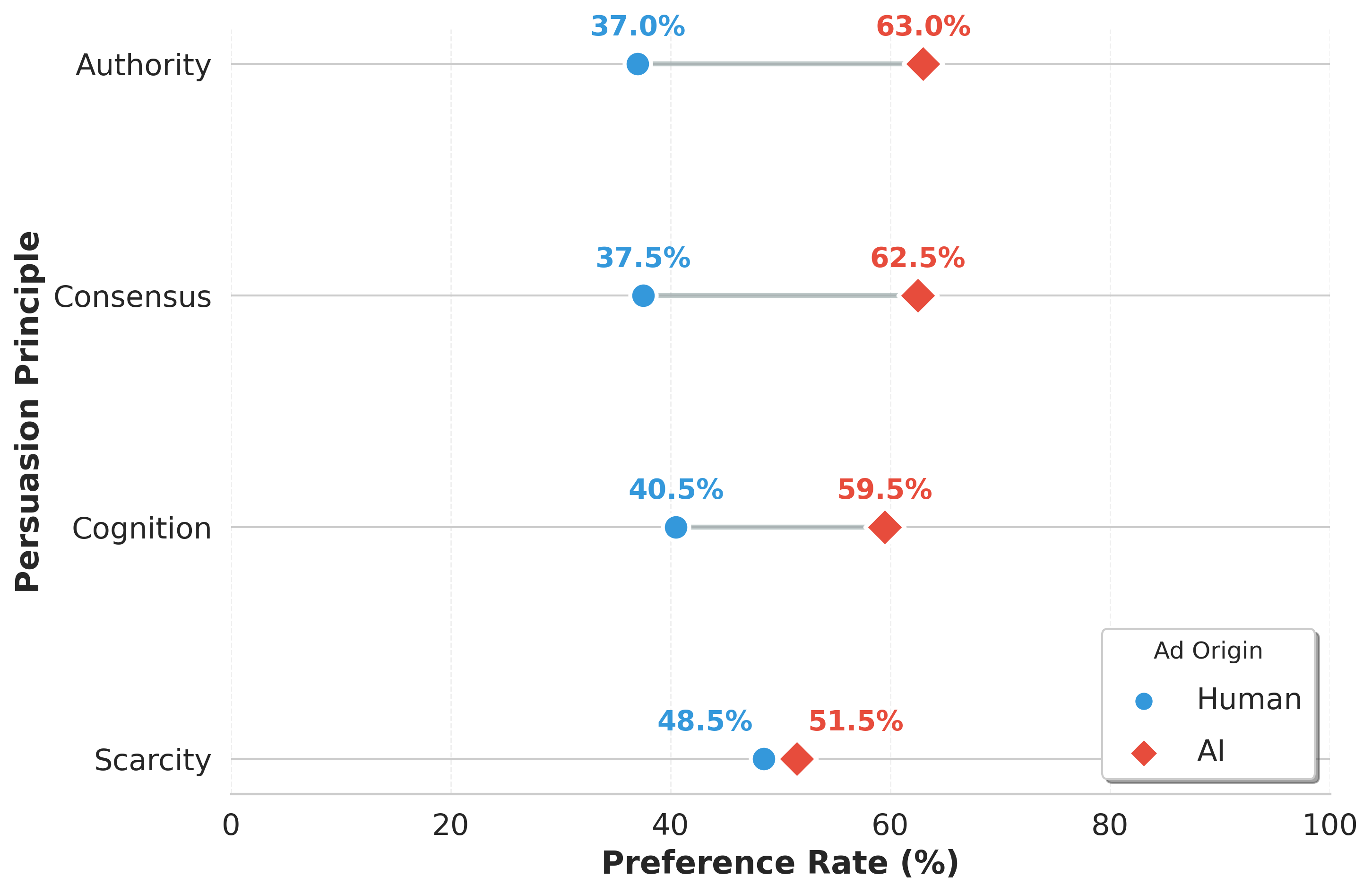}
        \caption{Performance by principle. AI holds a strong lead in \textbf{Authority} and \textbf{Consensus}, but ties on \textbf{Scarcity} due to tactical skepticism.}
        \label{fig:persuasion_dumbbell}
    \end{minipage}
\end{figure}

\subsection{Is AI-Generated Content Superior to Human Creativity?}
In the domain of persuasion principles (Study 2), AI did not merely compete with humans; it significantly outperformed them. With an 18.2 percentage point advantage overall, and dominant performance in \textit{Authority} (+26\%) and \textit{Consensus} (+25\%), the data suggests that AI models have internalized the structural rules of persuasion better than the average human creator.
As shown in Figure~\ref{fig:qualitative_themes}, qualitative analysis reveals that this superiority stems from \textbf{Message Sophistication} and \textbf{Visual Coherence}. Participants frequently described AI ads as ``inspiring,'' ``elevating,'' and ``professional''. This supports the theory of \textit{Narrative Transportation} \cite{green2000transportation}—the AI was better at telling a transformative story about the product rather than just describing it. Furthermore, the AI's ability to generate images that perfectly matched the textual ``vibe'' indicates that GenAI reduces the friction between copy and art direction. This creates a high degree of \textbf{Processing Fluency} \cite{alter2009uniting}, where the visual-textual alignment reduces cognitive load and acts as a heuristic cue for credibility.
However, this advantage was non-existent in the \textbf{Scarcity} condition, where AI and human ads performed equally (Figure~\ref{fig:persuasion_dumbbell}). This exception can be explained by the \textbf{Persuasion Knowledge Model (PKM)} \cite{friestad1994persuasion}. Scarcity tactics—such as ``Act Now!'' or ``Limited Time Only''—are distinctively overt attempts to curtail the consumer's freedom to choose. Unlike the subtle cues of Authority, Scarcity acts as a \textbf{persuasion trigger}: it alerts the user that they are being marketed to, causing them to raise a ``psychological shield'' of skepticism. Once this resistance is activated, the execution quality of the ad becomes irrelevant; the user resists the \textit{tactic} itself, not the creator. Thus, while AI can polish the aesthetics of urgency, it cannot bypass the structural resistance consumers have developed toward high-pressure sales techniques.

\begin{figure}[H]
    \centering
    \begin{subfigure}[b]{0.48\linewidth}
        \centering
        \includegraphics[width=\linewidth]{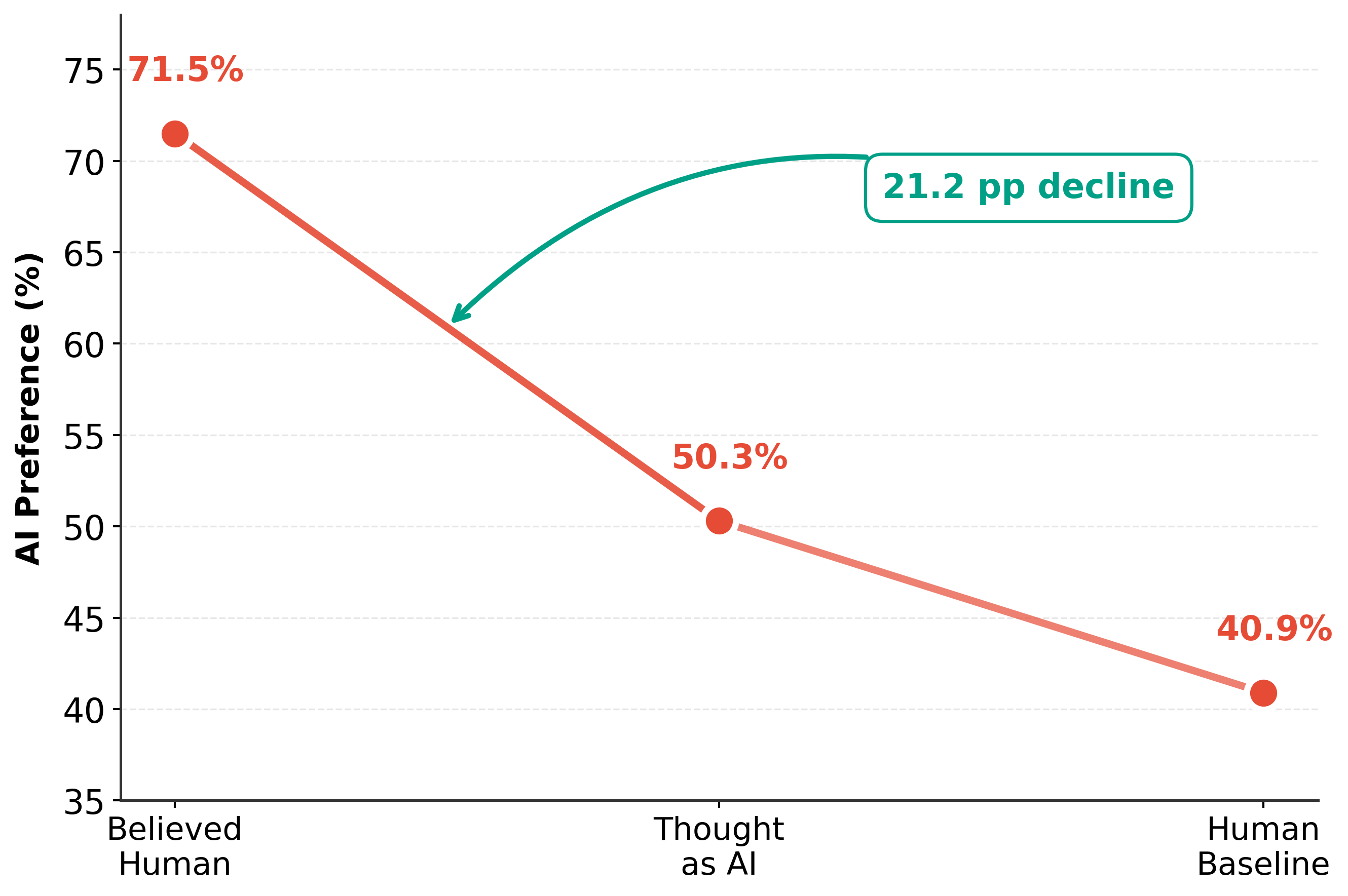}
        \caption{The Perception Paradox}
        \label{fig:paradox}
    \end{subfigure}
    \hfill
    \begin{subfigure}[b]{0.48\linewidth}
        \centering
        \includegraphics[width=\linewidth]{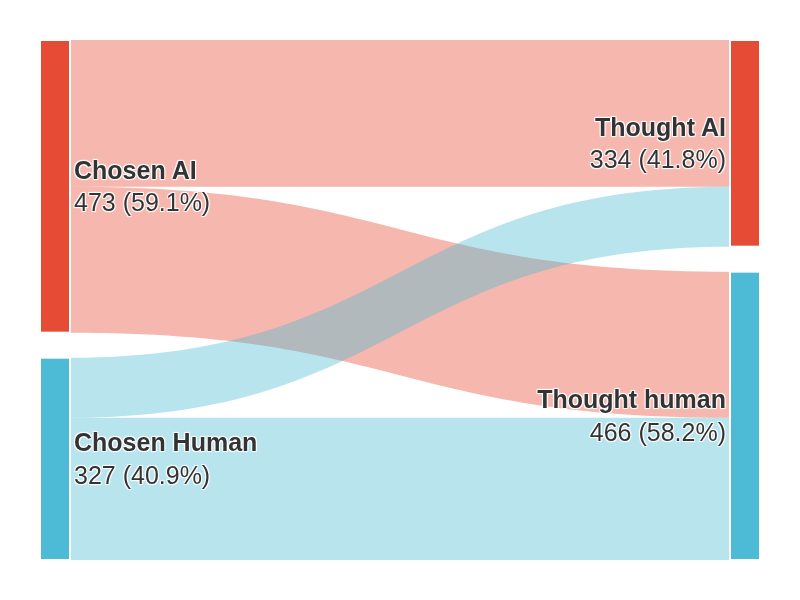}
        \caption{Choice-Perception Flow}
        \label{fig:sankey}
    \end{subfigure}
    
    \caption{Interaction between detection and preference. (a) AI preference drops when detected (21.2 pp penalty) but remains above the human baseline (40.9\%). (b) Attribution flow showing that AI preference is driven by both those who were deceived and those who accepted the AI origin.}
    \label{fig:combined_analysis}
\end{figure}

\subsection{Does the ``AI Perception'' hinder User Trust?}
A central concern in HCI is whether the revelation of non-human authorship breaks the social contract of communication. Our data confirms the existence of a \textbf{Detection Penalty}; correctly identifying an ad as AI-generated reduced preference by 21.2\% as shown in Figure~\ref{fig:combined_analysis}. This validates the ``authenticity requirement'' often discussed in marketing literature.
However, a paradox emerged: despite this penalty, the AI advertisements \textit{still} won. Even when participants knew they were looking at AI, 50.3\% preferred it over the human alternative. This phenomenon, which we term \textbf{Quality Resilience}, suggests that the functional superiority of the content overrides the negative bias associated with its origin. This hierarchy of effects can be mapped to the \textbf{Elaboration Likelihood Model (ELM)} \cite{petty1986elaboration}, while the ``source origin'' (AI) acts as a negative \textit{peripheral cue}, the superior ``argument quality'' (aesthetic appeal, aspirational messaging) serves as a dominant \textit{central cue} that drives the final decision. For nearly half of the participants, the decision was origin-independent; they prioritized the ``what'' over the ``who,'' signaling a shift toward a ``Post-Authenticity'' consumption model.

\subsection{Is Authenticity Still a Human Trait?}

Perhaps the most surprising finding is the generational divergence in how authenticity is perceived and detected (Figure~\ref{fig:age_detection}).

\textbf{Boomers (The Deceived):} Older participants (65+), who placed the highest value on ``genuine'' and ``natural'' qualities, frequently misidentified AI ads as human (Figure~\ref{fig:age_detection}). Qualitative analysis suggests they engaged in a heuristic \textbf{Halo Effect} \cite{nisbett1977telling}: they conflated high production quality with human effort. They attributed the AI's high-quality polish to ``human thoughtfulness,'' effectively using quality as a proxy for humanity and failing to spot the machine because it was ``too good'' to look fake.

\textbf{Gen Z (The Pragmatists):} Conversely, younger participants (18-24) were significantly more adept at spotting AI (73.0\% accuracy) but judged ads based on ``vibe'' rather than authenticity. This behavior aligns with \textbf{Uses and Gratifications Theory} \cite{katz1973uses}: these users prioritize the immediate gratification of the aesthetic experience over the provenance of the source. The ``uncanny valley'' appears to be closing for this cohort, not because they are deceived, but because they have learned to detach the artifact's utility from its creator.

\begin{figure}[t!]
    \centering
    \includegraphics[width=0.7\linewidth]{attachments/age_detection.png}
    \caption{Generational divergence in detection. Accuracy declines sharply with age, with digital natives (18-24) achieving 73.0\% accuracy while the 65+ cohort performs at chance level (48.3\%).}
    \label{fig:age_detection}
\end{figure}

% \begin{wrapfigure}{r}{0.5\textwidth} % 'r' = right alignment, '0.5\textwidth' = half page width
%     \centering
%     \vspace{-10pt} % Optional: Pulls the image up slightly to save header space
%     \includegraphics[width=\linewidth]{attachments/age_detection.png}
%     \caption{Generational divergence in detection. Accuracy declines sharply with age, with digital natives (18-24) achieving 73.0\% accuracy while the 65+ cohort performs at chance level (48.3\%).}
%     \label{fig:age_detection}
%     \vspace{-5pt} % Optional: Pulls the text below closer
% \end{wrapfigure}

\subsection{Limitations and Future Work}

Our studies have some limitations that open directions for future research. Both samples were drawn from online platforms in the United States, which limits cultural and regional generalizability. Real-world advertising effects depend on repeated exposure, competing products, and actual purchasing behavior, while our measures focused on single exposure preferences and self-reported intentions.

Methodologically, Study~1 examined only two personality traits and a single product category with short textual messages and a fixed image. Other traits or product types may interact differently with LLM-generated personalization. Study~2 explored four persuasion principles, but persuasion theory includes many additional mechanisms that may show distinct patterns when instantiated by AI. Our prompts and model choices were also conservative. We did not fine-tune models, optimize prompts iteratively, or dynamically adapt content based on user feedback, which suggests that our results are likely a lower bound on what such systems could achieve.

Future work can extend this line of research by testing multi-trait personalization, richer interaction histories, and longitudinal outcomes such as trust, brand attitude, and perceived manipulation. It will also be important to explore normative questions, such as when personality-based and principle-based personalization crosses the line from beneficial tailoring to exploitative targeting. Our findings indicate that, at least in controlled settings, AI-generated persuasion can already compete with human creativity. The central question is no longer whether AI can persuade, but how we choose to constrain and govern that capability.

\section{CONCLUSION}
\label{sec:conclusion}

This research investigated the capability of Large Language Models and diffusion models to generate effective advertising content, comparing AI-generated assets directly against human-authored counterparts across personality-based and principle-based frameworks. Our findings demonstrate that these models have reached a threshold of creative utility where they can effectively rival, and in specific contexts surpass, human expertise in ad creation.

In the domain of trait-based personalization, we observed that LLMs successfully generated engaging content for individuals high in openness, yielding performance metrics statistically equivalent to human writers. However, the efficacy of this matching is not universal; targeting neuroticism proved ineffective for both AI and human creators. This highlights the inherent challenge of engaging \textbf{prevention-focused} traits, where tailored stimuli are more likely to trigger avoidance behaviors than engagement, regardless of the author's origin.

When expanding the scope to multimodal advertisements based on persuasion principles, AI-generated content demonstrated a clear performance advantage, achieving a 59.1\% preference over human-created advertisements. This superiority was most pronounced in advertisements leveraging Authority and Consensus principles, suggesting that GenAI excels at synthesizing markers of credibility and social proof to establish high \textit{processing fluency}. However, this advantage was non-existent in the Scarcity condition, indicating that AI cannot inherently overcome consumer skepticism toward specific high-pressure tactics.

With reference to consumer perception, our results reveal a complex relationship between \textit{detection} and \textit{preference}. While participants possessed only a moderate ability to identify AI-generated content (58.4\% accuracy), successful detection resulted in a significant penalty, reducing the likelihood of preference by 21.2 percentage points. Yet, a critical paradox emerged: even when correctly identified as artificial, AI advertisements maintained a majority preference share (50.3\%) against human benchmarks. This suggests that the execution quality of the content—its visual coherence and message sophistication—possesses a \textbf{Quality Resilience} that can override the negative bias associated with non-human authorship.

These findings signal a shift toward a ``post-authenticity'' advertising era, characterized by a distinct generational divergence. Our demographic analysis revealed that while younger participants acted as \textbf{pragmatists}, identifying AI but prioritizing aesthetic appeal over origin, older participants frequently conflated high production quality with human authorship. As the younger cohort defines the future market, the distinction between human and machine authorship is becoming secondary to the quality of the experience, now that AI has demonstrated it can rival human creative proficiency. The evidence suggests that GenAI has evolved beyond a simple productivity aid, establishing itself as a sophisticated engine for production capable of delivering superior engagement outcomes.

Ultimately, this work documents a pivotal moment in the evolution of persuasive communication. The question is no longer whether artificial intelligence can match human creativity in advertising—it can, and in measurable ways, it already exceeds it. The questions that remain are societal: how do we ensure that these powerful capabilities serve consumer interests rather than exploit psychological vulnerabilities? How do we balance personalization benefits against privacy concerns and manipulation risks? And as the line between human and machine-generated content continues to blur, what new frameworks for authenticity, trust, and creative attribution must we construct? These are the challenges that will define the next chapter of AI-driven persuasion.

% \begin{acks}
% This work is partially supported by the Australian Research Council (ARC) Training Centre for Information Resilience (Grant No. IC200100022).
% \end{acks}

\bibliographystyle{ACM-Reference-Format}
\bibliography{persuasionLitRev}

\clearpage % Forces a new page and ensures previous figures are printed
\appendix

\section{Appendix: Qualitative Analysis of Participant Preferences}
\label{sec:appendix_qualitative}

\subsection{Thematic Analysis Methodology}
To understand the underlying drivers of advertisement preference, we analyzed open-ended responses from participants across all four persuasion conditions. Responses were categorized using an inductive thematic analysis approach, focusing on three primary dimensions: message content, visual aesthetics, and perceived authenticity. This process yielded three dominant drivers of AI success: Message Sophistication (mentioned in 38\% of responses), Visual Coherence (32\%), and Psychological Precision.

\subsection{Drivers of AI Superiority}
The most significant differentiator for AI-generated content was the use of \textit{aspirational messaging}. While human-authored advertisements were frequently described as ``relatable,'' ``simple,'' and ``direct,'' AI-generated narratives were characterized by participants as ``inspiring'' (36 mentions), ``elevating'' (15 mentions), and ``professional'' (44 mentions). This suggests that the LLM successfully employed \textit{narrative transportation}, moving beyond functional product descriptions to offer transformative storylines that resonated with users seeking improvement or status.

In terms of \textit{visual coherence}, participants consistently noted that AI-generated imagery achieved superior integration with the textual message. Comments highlighted that AI advertisements possessed better ``color palettes'' and ``contrasts,'' with one participant noting the imagery set the mood of a ``professor's lecture room,'' thereby reinforcing the Authority principle. Conversely, human-created visuals were often criticized for lacking this cohesion, though they were praised for appearing more ``natural'' and ``pensive.''

\begin{table*}[h]
\caption{Comparative Qualitative Feedback: Representative participant quotes contrasting AI and Human advertisements.}
\label{tab:qualitative_quotes}
\centering
\begin{tabular}{p{0.45\textwidth} | p{0.45\textwidth}}
\hline
\textbf{Participants who Chose AI} & \textbf{Participants who Chose Human} \\
\hline
\multicolumn{2}{c}{\textit{On Message Sophistication}} \\
\hline
``It speaks to how the product will improve your life, not just who it's made for.'' & ``Felt more relatable and approachable. The tagline is simple, direct.'' \\
``The captivating tagline and the imagery is splendid and precise.'' & ``More of a marketing slogan than a personal message.'' \\
``More influencing than A which looks like demanding.'' & ``Too abstract and formal.'' \\
\hline
\multicolumn{2}{c}{\textit{On Visual Presentation}} \\
\hline
``The color palette. The colors contrast better.'' & ``The thoughtful pose felt natural and inviting.'' \\
``The image is more fitting with the text... the overall darker colors give it the mood of a professor's lecture room.'' & ``I like the contrast of colors. The image on the right has too much brown.'' \\
``It is so clear and realistic.'' & ``Looks more pensive.'' \\
\hline
\end{tabular}
\end{table*}

\subsection{The Perception Paradox}
Our analysis revealed a complex interaction between detection and preference. As detailed in Table \ref{tab:perception_groups}, the largest sub-group of AI choosers (50.3\%) believed the content was human-created, attributing ``genuine thoughtfulness'' and ``real human creativity'' to the AI output. However, a substantial group (49.7\% of AI choosers) selected the AI advertisement despite correctly identifying its origin. This indicates a shift toward \textit{Quality Resilience}, where the sophisticated execution of the AI content overrides the ``authenticity penalty'' usually associated with automated media.

\begin{table}[h]
\caption{Perception Groups: Interaction between Origin Identification and Advertisement Choice.}
\label{tab:perception_groups}
\centering
\begin{tabular}{l c c p{4cm}}
\hline
\textbf{Group} & \textbf{Count} & \textbf{\% of Total} & \textbf{Key Insight} \\
\hline
Chose AI, Detected AI & 235 & 29.4\% & \textbf{Quality Resilience:} Embraced sophistication despite knowing origin. \\
Chose AI, Believed Human & 238 & 29.8\% & \textbf{The Deceived:} Attributed ``genuine thoughtfulness'' to AI work. \\
Chose Human, Detected Human & 232 & 29.0\% & \textbf{Authenticity Seekers:} Valued ``natural'' qualities. \\
Chose Human, Believed AI & 95 & 11.9\% & \textbf{Avoidance:} Selected perceived lower-quality content to avoid AI. \\
\hline
\end{tabular}
\end{table}

\end{document}